\documentclass[prd,onecolumn]{revtex4}
\usepackage{dcolumn}
\usepackage{multirow}
\usepackage{graphicx}
\usepackage{amssymb}
\usepackage{bm}
\usepackage{hyperref}%for .pdf
\usepackage{epstopdf}%for .pdf
\usepackage{color}
\usepackage{mathrsfs}
\usepackage{amsmath,amssymb,amsthm}
\usepackage{rotating}
\usepackage{sverb, longtable}
\usepackage{subfigure}

\usepackage[graphicx]{realboxes}
\usepackage{adjustbox}
\begin{document}

\title{Novel Physics with International Pulsar Timing Array: Axionlike Particles, Domain Walls and Cosmic Strings}
\author{Deng Wang}
\email{cstar@nao.cas.cn}
\affiliation{National Astronomical Observatories, Chinese Academy of Sciences, Beijing, 100012, China}
\begin{abstract}
After NANOGrav, the IPTA collaboration also reports a strong evidence of a stochastic gravitation wave background. This hint has very important implications for fundamental physics.
With the recent IPTA data release two, we attempt to search signals of light new physics. 
and give new constraints on the audible axion, domain walls and cosmic strings models. We find that the best fit point corresponding to a decay constant $F\approx5\times10^{17}$ GeV and an axion mass $m_a\approx2\times10^{-13}$ eV from NANOGrav data is ruled out by IPTA at beyond $2\sigma$ confidence level. Fixing the coupling strength $\lambda=1$, we obtain a $2\sigma$ lower bound on the breaking scale of $Z_2$ symmetry $\eta>135$ TeV. Interestingly, we give a very strong restriction on the cosmic-string tension
$\mathrm{log}_{10}\,G\mu=-8.93_{-0.06}^{+0.12}$ at $1\sigma$ confidence level. Employing the rule of Bayes factor, we find that IPTA data has a moderate, strong and inconclusive preference of an uncorrelated common power-law (CPL) model over audible axion, domain walls and cosmic strings, respectively. This means that it is hard to distinguish CPL from cosmic strings with current observations and more pulsar timing data with high precision are required to give new clues of underlying physics.

\end{abstract}
\maketitle

\section{Introduction}
The discovery of gravitational waves (GWs) from a binary black holes merger by the LIGO collaboration \cite{LIGOScientific:2016aoc} has opened a new window to study the evolution of the universe, and prompted human beings to step into a new era of gravitational wave (GW) astronomy. Since LIGO's discovery, various detectors which detect different frequencies of GWs has been proposed and developed. As is well known, LIGO can detect the compact binary mergers in the frequency range $10-10^3$ Hz. For the purpose of detecting the low frequency GW sources such as massive binaries and supernovae, the space-based GW detectors such as eLISA \cite{Klein:2015hvg} have been proposed, which is designed to operate in the frequency range $10^{-5}-1$ Hz. In addition, pulsar timing arrays (PTA) \cite{Manchester:2013ndt} and SKA \cite{Dewdney2009} are aimed at probing the stochastic gravitational wave backgrounds (SGWB) around the very low frequency $10^{-9}$ Hz. All the above experiments will help us understand the universe better. 

Since GWs are hardly disturbed during their travels through cosmic spacetime, they can carry the information of the early universe before the CMB epoch. Recently, it is exciting that NANOGrav \cite{Brazier:2019mmu}, PPTA \cite{Kerr:2020qdo}, EPTA \cite{Desvignes:2016yex} and IPTA \cite{Perera:2019sca} have reported successively the strong evidence of a stochastic common spectrum process at low frequencies, although PPTA group prefers discreetly identifying their result as an unknown systematic uncertainty. Nonetheless, there is no evidence found for a spatial correlation predicted by general relativity. Such a stochastic GW background can be explained in the early universe by various physical processes, e.g., phase transitions \cite{Kosowsky:1992rz, Caprini:2010xv,Nakai:2020oit,Addazi:2020zcj,Ratzinger:2020koh,Li:2021qer}, axionlike particles \cite{Ratzinger:2020koh,Machado:2018nqk, Machado:2019xuc,Salehian:2020dsf}, domain walls \cite{Hiramatsu:2013qaa,Kadota:2015dza}, cosmic strings \cite{Siemens:2006yp, Blanco-Pillado:2017rnf,Ellis:2020ena,Blasi:2020mfx} and primordial black hole formation \cite{Vaskonen:2020lbd,DeLuca:2020agl,Kohri:2020qqd}. In practice, giving accurate constraints on these sources and distinguishing them efficiently via observations is an important task. In light of the recent IPTA data release two (DR2) \cite{Perera:2019sca} which consists of 65 pulsars, we are motivated by exploring the signals of light new physics and give new constraints on the audible axion, domain walls and cosmic strings.   

This work is outlined in the following manners. In the next section, we introduce briefly three SGWB models. In section III, we carry out the numerical analysis and exhibit the results. The discussions and conclusions are presented in the final section. 

\section{Models}
We will introduce briefly three SGWB models including axionlike particles, domain walls and cosmic strings.

\subsection{Axionlike particles}
The audible axion model is fistly proposed in Ref.\cite{Machado:2019xuc}, which consists of an axion field $\phi$ and a massless dark photon $X_\mu$ of an unbroken $U(1)_X$ Abelian gauge group,
\begin{equation}
\frac{\mathcal{L}}{\sqrt{-g}}=\frac{1}{2}\partial_\mu\phi\partial^\nu\phi-V(\phi)-\frac{1}{4}X_{\mu\nu}X^{\mu\nu}-\frac{q}{4F}\phi X_{\mu\nu}\tilde{X}^{\mu\nu},     \label{1}
\end{equation}
where $F$ denotes the axion decay constant, i.e., the scale where the global symmetry corresponding to the Nambu-Goldstone field $\phi$ is broken and produces the light pseudoscalar $\phi$, $q$ is a dimensionless charge, $X_{\mu\nu}$ and $\tilde{X}_{\mu\nu}$ represent the dark photon field strength tensor and its dual, and the axion potential $V(\phi)=m_a^2F^2[1-\cos(\phi/F)]$ where $m_a$ is the axion mass.  

In the axion misalignment mechanism, we use the traditional assumption that the axion is perturbed and displaced from the minimum of its potential $V(\phi)$ by $\theta F$ with $\theta\sim \mathcal{O}(1)$, after the inflation ends. Until the cosmic expansion rate is of the same order as $m_a$, the axion stops being displaced and starts to oscillate around the origin. When the axion rolls in the early universe, it is possible to produce efficient energy transfer to dark photons due to rolling induced tachyonic instability. This process amplifies the quantum fluctuations in the dark photon field, which evolves over time and forms the detectable SGWB at macroscopic scales today.  

The GW spectrum generated by audible axions is very closely related to the axion mass, and has a peak at the frequency where the dark photon momentum mode grows fastest. Following Ref.\cite{Ratzinger:2020koh}, the strength of the axion source, namely the energy from axion, determines the GW amplitude in this model. To a large extent, this amplitude will be affected by the axion decay constant $F$. The present peak amplitude of this GW signal is roughly expressed as
\begin{equation}
\Omega_{\mathrm{GW}}h^2\approx 1.84\times 10^{-7}\left(\frac{50\,\theta^2}{q}\right)^{\frac{4}{3}}\left(\frac{F}{m_{pl}}\right)^4, \label{2}
\end{equation}
where $h\equiv H_0/(100$ km s$^{-1}$ Mpc$^{-1}$) denotes the dimensionless Hubble parameter and $m_{pl}$ is the Plank mass. Today's peak frequency of GW spectrum is approximated as 
\begin{equation}
f_\mathrm{p}^0\simeq 1.1\times 10^{-8}\left(\frac{m_a}{10^{-15}\, \mathrm{eV}}\right)^{\frac{1}{2}}\left(\frac{q\theta}{50}\right)^{\frac{2}{3}} \mathrm{Hz}. \label{3}
\end{equation}
Furthermore, in order to implement numerical computations, we take the GW spectrum specified in Ref.\cite{Ratzinger:2020koh}
\begin{equation}
\Omega_{\mathrm{GW}}(f)h^2=\frac{6.3\,\Omega_{\mathrm{GW}}^0h^2\left(\frac{f}{2f_\mathrm{p}^0}\right)^{\frac{3}{2}}}{1+\left(\frac{f}{2f_\mathrm{p}^0}\right)^{\frac{3}{2}}\mathrm{exp}\left(\frac{12.9f}{2f_\mathrm{p}^0}-1\right)}, \label{4}
\end{equation}
and set $\theta=1$ and $q=50$.

\subsection{Domain walls}
Domains walls are sheet-like objects formed in the early universe when a discrete symmetry is spontaneously broken \cite{Hiramatsu:2013qaa}. As is well known, stable domain walls existing in the universe are inconsistent with the standard cosmology, because their energy density tends to be dominated in the total cosmic energy budget. Nonetheless, unstable domain walls which annihilate at sufficiently early times and do not affect the evolution of the universe can exist. They can act as the cosmological source of a SGWB. 

In this work, we consider a real scalar field model. Its Lagrangian density reads as \cite{Hiramatsu:2013qaa}
\begin{equation}
\mathcal{L}=\frac{1}{2}\partial_\mu\partial^{\mu}\phi-V(\phi), \label{5}
\end{equation}
with a double well potential 
\begin{equation}
V(\phi)=\frac{\lambda}{4}(\eta^2-\phi^2)^2, \label{6}
\end{equation}
Where the coupling strength $\lambda$ and the breaking scale of $Z_2$ symmetry $\eta$ are two free parameters. After adding the correction term $\lambda T^2\phi^2/8$ to the above potential in the early universe with a finite temperate $T$, the discrete $Z_2$ symmetry ($\phi\rightarrow-\phi$) is recovered. When the temperature of the universe decreases with its expansion and is smaller than the critical value $T_c=2\eta$, $Z_2$ symmetry is spontaneously broken to form domain walls. After domains walls are formed, due to their surface tension, their curvature radius is fast homogenized. They will evolve to the so-called scaling regime, where typical scales of the network consisting of them such as curvature radius and distance between neighboring walls will be comparable to the Hubble radius \cite{Press:1989yh,Garagounis:2002kt,Leite:2011sc,Leite:2012vn}.  

In this scaling regime, domain walls lose their energy and maintain the scaling property by their self-interaction such as changing their shape or collapsing into closed walls. A part of energy of domain walls are released as GWs during this process. Therefore, domain walls decay can also serve as the cosmological source of a SGWB.

To perform numerical fits, following Ref.\cite{Hiramatsu:2013qaa}, we show the present GW peak amplitude as 
\begin{equation}
\Omega_{\mathrm{GW}}h^2\approx 1.0\times10^{-21}\lambda^2\epsilon^{-2}\left(\frac{g_\star}{100}\right)^{-\frac{1}{3}}\left(\frac{\eta}{10^{15}\,\mathrm{GeV}}\right)^4, \label{7}
\end{equation}
where $\epsilon$ is a free parameter, and express the peak frequency of GW spectrum as
\begin{equation}
f_\mathrm{p}^0\simeq 6.7\times 10^{9}\lambda^{-\frac{1}{4}}\epsilon^{\frac{1}{2}}\left(\frac{\eta}{10^{15}\,\mathrm{GeV}}\right)^{\frac{1}{2}} \mathrm{Hz}. \label{8}
\end{equation}
We will fix $\lambda=1$ and $g_\star=100$ during the numerical calculations, and 
consider the frequency dependence $\Omega_{\mathrm{GW}}\propto f^3$ for $f<f_\mathrm{p}^0$ and $\Omega_{\mathrm{GW}}\propto f^{-1}$ for $f>f_\mathrm{p}^0$ for the domain walls model \cite{Hiramatsu:2013qaa}.

\subsection{Cosmic strings}
Many models with new physics beyond the Standard Model of particle physics predict phase transitions \cite{Mazumdar19}, which lead to the spontaneous breaking of $U(1)$ symmetry in the early universe. A common prediction is that these phase transitions will phenomenologically generate a network of cosmic strings \cite{Kibble1976,Jeannerot:2003qv}, which are 1-dimensional stable objects characterized by their typical tension $\mu$. Cosmic strings can form loops which release energy and shrink by emitting GWs. Hence, they can also serve as the cosmological source of a SGWB. Since the primordial GW signal from a network of cosmic strings contains crucial information about ultraviolet physics, it is important to search for such a signal with current and future GW experiments across a vast range of GW frequencies.

In this study, we take the widely used Nambu-Goto cosmic string model and use a simple method to compute the GW spectrum from a network of cosmic strings. The GW spectrum can be written as 
\begin{equation}
\Omega_{\mathrm{GW}}(f)=\sum\limits_{k=1}^\infty k \Gamma^{(k)}\Omega_{\mathrm{GW}}^{(k)}(f), \label{9}
\end{equation}
where we adopt the total emission rate $\Gamma\approx50$ in order to be compatible with numerical simulations \cite{Blanco-Pillado:2013qja,Vilenkin1981,Turok1984,Quashnock1990}, and we also assume that GWs released by a network of cosmic strings are dominated by cusps propagating along cosmic-string loops with $\Gamma^{(k)}=\Gamma k^{-(4/3)}/(\Sigma^\infty_{m=1}m^{-(4/3)})$. The contribution of each mode in Eq.(\ref{9}) is shown as 
\begin{equation}
\Omega_{\mathrm{GW}}^{(k)}(f)=\frac{16\pi(0.1)(G\mu)^2}{3H_0^2\alpha(\alpha+\Gamma G\mu)f}\int_{t_F}^{t_0}d\tilde{t}\,\frac{C_{\mathrm{eff}}(t_i)}{t_i^4}\left[\frac{a(t_i)}{a(\tilde{t})}\right]^3\left[\frac{a(\tilde{t})}{a(t_0)}\right]^5\Theta(t_i-t_F), \label{10}
\end{equation}
where $G\mu$ denotes string tension, $\alpha$ is initial loop size, $a$ is scale factor, $t_F$ is the network formation time, $\tilde{t}$ is GW emission time, $t_0$ is current time, $C_{\mathrm{eff}}$ controlling the string loop number density is 5.4 (0.39) \cite{Cui:2017ufi,Cui:2018rwi} in the radiation (matter) dominated era, and the factor 0.1 comes from numerical simulations \cite{Blanco-Pillado:2019vcs,Blanco-Pillado:2019tbi}, which suggests only this fraction of energy can produce large string loops and then release GWs efficiently. String loops emit at normal oscillation mode frequencies, letting us to show the frequency corresponding to the $k$-th mode as 
\begin{equation}
f=\frac{a(\tilde{t})}{a(t_0)}\frac{2k}{\alpha t_i+\Gamma G\mu(t_i-\tilde{t})}. \label{11}
\end{equation}
By combing Eqs.(\ref{9}-\ref{11}), one can easily derive the GW energy density spectrum $\Omega_{\mathrm{GW}}(f)h^2$ of cosmic strings.

\section{Analysis and results}

\begin{figure}
	\centering
	\includegraphics[scale=0.5]{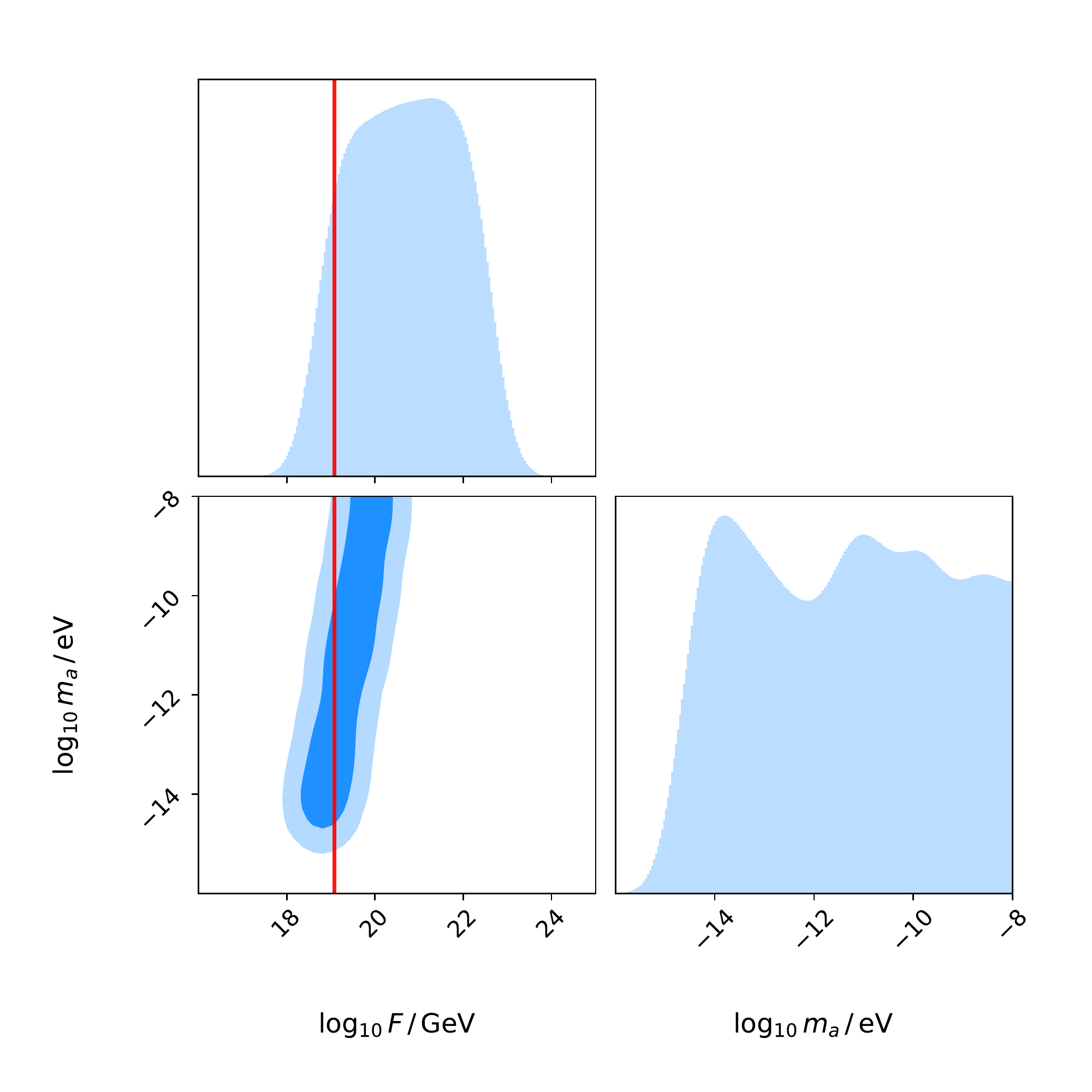}
	\caption{The marginalized posterior distributions of free parameters in the audible axion model. The vertical lines denote the Planck mass $m_{pl}=1.2\times10^{19}$ GeV.} \label{f1}
\end{figure}

\begin{figure}
	\centering
	\includegraphics[scale=0.5]{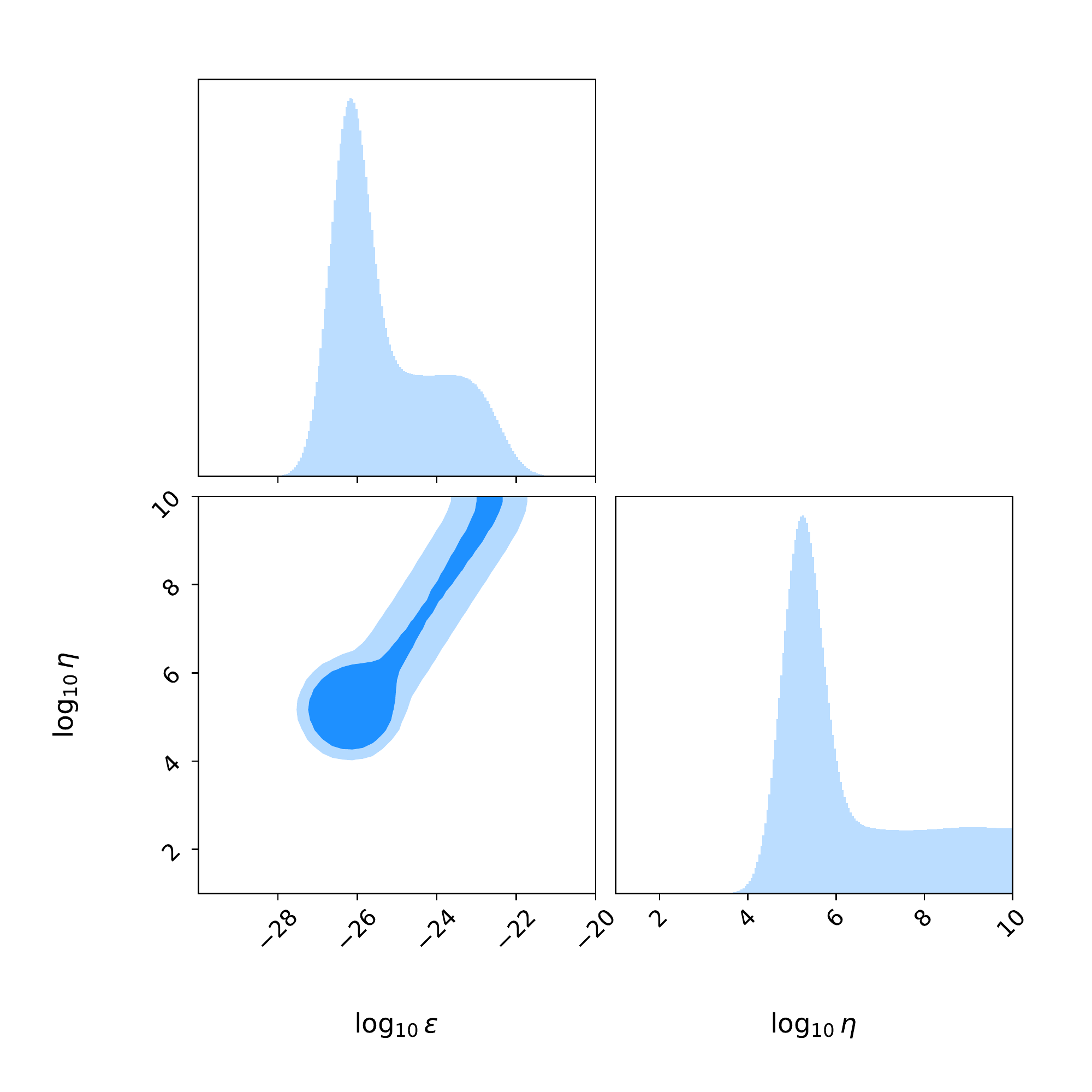}
	\caption{The marginalized posterior distributions of free parameters in the domain wall decay model.} \label{f2}
\end{figure}

\begin{figure}
	\centering
	\includegraphics[scale=0.5]{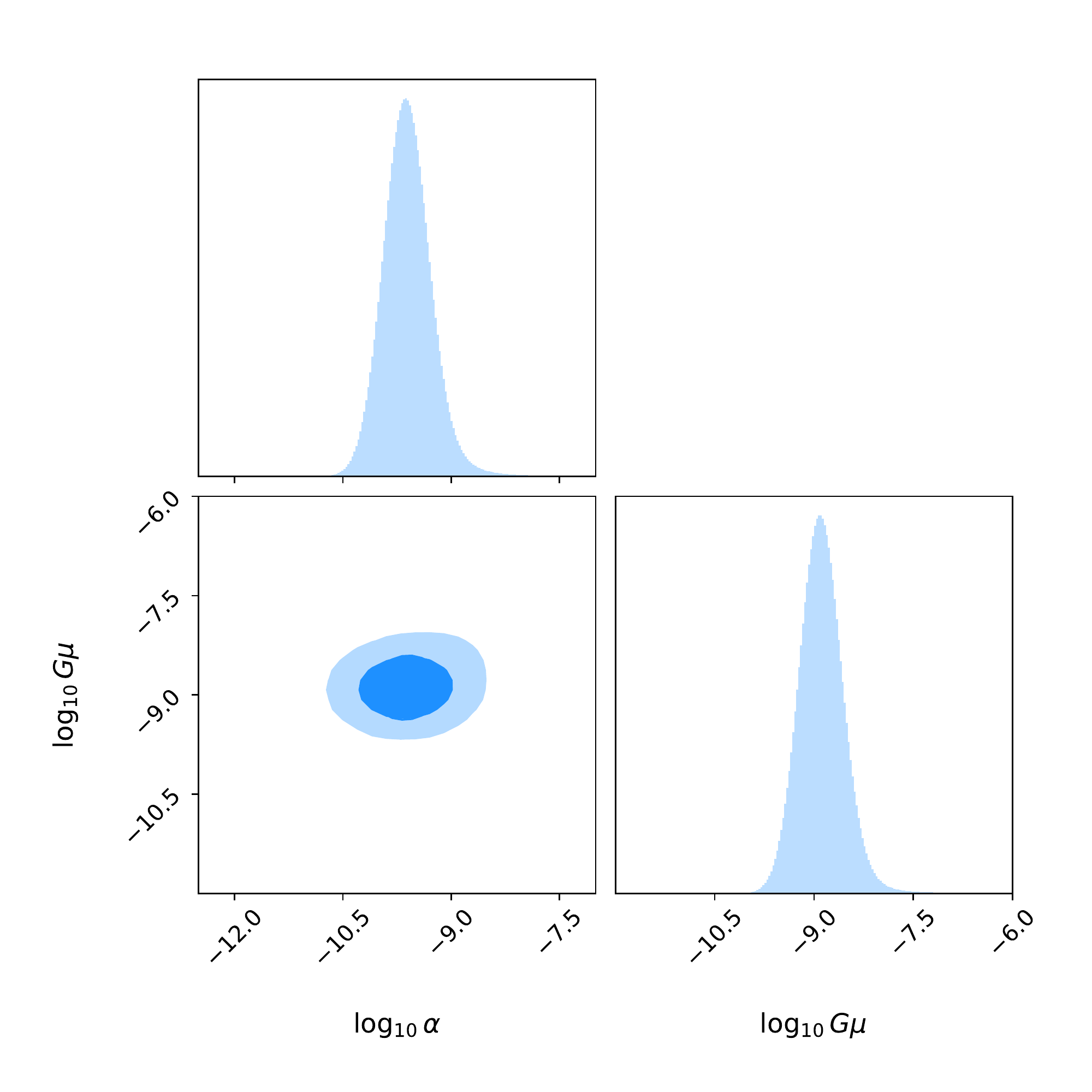}
	\caption{The marginalized posterior distributions of free parameters in the cosmic string model.} \label{f3}
\end{figure}

With the recent IPTA DR2 \cite{Perera:2019sca}, which consists of 65 pulsars, we are dedicated to explore underlying new physics. Specifically, we employ the IPTA DR2 posterior distributions on the delay spectrum \cite{IPTADR2} as input data and then perform the Marcov chain Monte Carlo analysis.
The marginalized posterior distributions of free parameters and constraining results for three considered SGWB models are shown in Figs.\ref{f1}-\ref{f3} and Tab.\ref{t1}, respectively. 

In Ref.\cite{Ratzinger:2020koh}, the parameter space has been constrained via NANOGrav 12.5-year data and the corresponding best fit point is a decay constant $F\approx5\times10^{17}$ GeV and an axion mass $m_a\approx2\times10^{-13}$ eV. However, in Fig.\ref{f1} and Tab.\ref{f1}, we find this point has been ruled out by IPTA data at beyond $2\sigma$ confidence level. Current constraints are $\mathrm{log}_{10}\,F=20.72_{-1.35}^{+1.30}$ and $\mathrm{log}_{10}\,m_a>-14.16$. Different from NANOGrav, we can just obtain the $2\sigma$ lower bound on axion mass with IPTA.
Since the permitted parameter space must satisfy the condition that the decay constant $F$ should be smaller than the Planck mass $m_{pl}$, the large part of parameter space in Fig.\ref{f1} is excluded by IPTA.

\begin{table*}
	\renewcommand\arraystretch{1.5}
	\caption{The confidence ranges of free parameters and logarithmic Bayes factors for the audible axion, domain walls and cosmic strings models from IPTA DR2. To compute the Bayes factors, our reference model is CPL.}
	\setlength{\tabcolsep}{3mm}{
		{\begin{tabular}{@{}ccccccccc@{}} \toprule
				Parameters      &$\mathrm{log}_{10}F$      &$\mathrm{log}_{10}m_a$           & $\mathrm{log}_{10}\epsilon$                 &$\mathrm{log}_{10}\eta$     &$\mathrm{log}_{10}G\mu$ &$\mathrm{log}_{10}\alpha$   & $\mathrm{ln}\,B_{ij}$                 \\ \colrule
				Audible axion       &$20.72_{-1.35}^{+1.30}$  &  $>-14.16$ ($2\,\sigma$)    &---   &---   &---    &---         &-3.69          \\  
				Domain walls           &---  &---          &$-26.09_{-0.15}^{+2.45}$         
				&$>5.13$ ($2\,\sigma$)    &---       &---             &-6.97                        \\
				Cosmic strings                  &---       &---       &---   &---    &$-8.93_{-0.06}^{+0.12}$ &$-9.63_{-0.12}^{+0.16}$                     &-0.85                       \\
				\botrule
			\end{tabular}
			\label{t1}}}
\end{table*}

For the case of domain walls, we obtain a $2\sigma$ lower bound on the breaking scale of $Z_2$ symmetry $\eta>135$ TeV by fixing the coupling $\lambda=1$. During the process of statistical analysis, if we fix $\lambda$ and $g_\star$, $\epsilon$ actually characterizes the amplitude of GW spectrum and we get the constraint on this effective amplitude parameter $\mathrm{log}_{10}\,\epsilon=-26.09_{-0.15}^{+2.45}$. From Fig.\ref{f2}, we find that the main parameter space concentrate around the best fit ($\epsilon=8.1\times10^{-27}$, $\eta=186$ TeV). This 2-dimensional property can be easily deduced from 1-dimensional distributions of two parameters.

In Refs.\cite{Blanco-Pillado:2017rnf,Ellis:2020ena,Blasi:2020mfx}, cosmic strings as an underlying GW source have been confronted with NANOGrav 12.5 year data and relative loose constraint are obtained. In light of IPTA DR2 30-frequencies data, we obtain tight constraints on the loop size $\mathrm{log}_{10}\,\alpha=-9.63_{-0.12}^{+0.16}$ and cosmic-string tension $\mathrm{log}_{10}\,G\mu=-8.93_{-0.06}^{+0.12}$ at $1\sigma$ confidence level. Furthermore, according to the relation between the string tension $G\mu$ and the underlying energy scale $s$ of $U(1)$ symmetry breaking \cite{Hind1995}, $s\sim10^{19.5}(G\mu)^{0.5}$ GeV, we find that IPTA data supports the breaking scale $s\sim[1.01, \, 1.24]\times10^{15}$ GeV. This gives a very strong constraint on the $U(1)$ symmetry breaking scale and may imply a deep link between the IPTA signal and novel physics related to the grand unification \cite{King:2020hyd}. 

Besides confronting different models with new data, another important task is distinguishing them from each other. Here,
we calcualte the Bayesian evidence of each GW source model, $\varepsilon_i$ and Bayes factor, $B_{ij}=\varepsilon_i/\varepsilon_j$, where $\varepsilon_j$ is the evidence of reference model. We use the so-called Jeffreys’ scale \cite{Trotta:2005ar}, i.e., $\mathrm{ln}\,B_{ij}=0-1$, $1-2.5$, $2.5-5$ and $>5$ indicate an {\it inconclusive}, {\it weak,} {\it moderate} and {\it strong} preference of the model $i$ relative to reference model $j$. For an experiment that leads to
$\mathrm{ln}\,B_{ij}<0$, it means the reference model is preferred by
data. As in our previous work \cite{Wang:2022wwj}, we choose the CPL model as our reference model, whose characteristic strain is $h_c(f)=A_{\mathrm{CPL}}(f/f_{\mathrm{yr}})^{(3-\gamma_{\mathrm{CPL}})/2}$ in the frequency range $f\in(f_l, f_h)$, where $f_{yr}=1$ $\mathrm{yr}^{-1}$, and $A_{\mathrm{CPL}}$ and $\gamma_{\mathrm{CPL}}$ are amplitude and spectral slope. The Bayesian evidence value of CPL is -44.464. The corresponding values for other three models are presented in Tab.\ref{t1}. One can easily find that IPTA data has a strong preference of CPL over Domain walls and a moderate preference of audible axion over CPL, and that there is no statistical preference between cosmic strings and CPL models.

\section{Discussions and conclusions}
After NANOGrav firstly reported a strong evidence of stochastic gravitational wave background, IPTA recently also claimed the same conclusion but with more complete pulsar timing data. We are motivated by using this better dataset to constrain light new physics beyond the Standard Model. Specifically, we consider audible axion, domain walls and cosmic strings. The GW peak is very sensitive to the axion mass, $Z_2$ symmetry breaking scale or spontaneous $U(1)$ symmetry breaking scale. Hence, IPTA data can well probe the parameter spaces of these three models in the PTA range.   

For the audible axion model, we find that the best fit point corresponding to a decay constant $F\approx5\times10^{17}$ GeV and an axion mass $m_a\approx2\times10^{-13}$ eV supported by NANOGrav has been ruled out by IPTA at beyond $2\sigma$ confidence level. The remained parameter space in $F$-$m_a$ plane may be explored by future experiments such as CASPEr \cite{JacksonKimball:2017elr}. 

For domain walls, setting the coupling strength $\lambda=1$, we obtain a $2\sigma$ lower bound on the $Z_2$ symmetry breaking scale $\eta>135$ TeV. It is interesting that the main parameter space in $\epsilon$-$\eta$ plane concentrates around the best fit ($\epsilon=8.1\times10^{-27}$, $\eta=186$ TeV). 

For cosmic strings, different from NANOGrav, we obtain a very tight constraint on model parameters, i.e., the loop size $\mathrm{log}_{10}\,\alpha=-9.63_{-0.12}^{+0.16}$ and cosmic-string tension $\mathrm{log}_{10}\,G\mu=-8.93_{-0.06}^{+0.12}$ at $1\sigma$ confidence level. It is intriguing that IPTA data supports the spontaneous symmetry breaking scale $s\sim[1.01, \, 1.24]\times10^{15}$ GeV, which gives a very strong restriction on the $U(1)$ symmetry breaking scale and may indicate a deep connection between the IPTA signal and novel physics related to the theory of grand unification. 

Interestingly, via the Bayes factor, we find that current IPTA DR2 data has a moderate and strong preference of CPL over audible axion  and domain walls, respectively, and that it is hard to distinguish CPL from cosmic strings with current data. This may imply that the CPL model will stand fro a long time and more high precision pulsar timing data are needed to probe the life space of new physics.

\section*{Acknowledgements}
Deng Wang thanks Liang Gao, Jie Wang and Qi Guo for hepful discussions, and Yan Gong and Shi Shao for useful communications. This work is supported by the National Nature Science Foundation of China under Grants No.11988101 and No.11851301.


\begin{thebibliography}{99}
%\cite{LIGOScientific:2016aoc}
\bibitem{LIGOScientific:2016aoc}
B.~P.~Abbott \textit{et al.} [LIGO Scientific and Virgo],
``Observation of Gravitational Waves from a Binary Black Hole Merger,''
Phys. Rev. Lett. \textbf{116}, no.6, 061102 (2016).

%\cite{Klein:2015hvg}
\bibitem{Klein:2015hvg}
A.~Klein \textit{et al.},
``Science with the space-based interferometer eLISA: Supermassive black hole binaries,''
Phys. Rev. D \textbf{93}, no.2, 024003 (2016).

%\cite{Manchester:2013ndt}
\bibitem{Manchester:2013ndt}
R.~N.~Manchester,
``The International Pulsar Timing Array,''
Class. Quant. Grav. \textbf{30}, 224010 (2013).

%\cite{Dewdney2009}
\bibitem{Dewdney2009}
P. E. Dewdney {\it et al.},
``The Square Kilometre Array,''
IEEE Proc. \textbf{97}, 1472 (2009).

%\cite{Brazier:2019mmu}
\bibitem{Brazier:2019mmu}
A.~Brazier \textit{et al.},
``The NANOGrav Program for Gravitational Waves and Fundamental Physics,''
[arXiv:1908.05356 [astro-ph.IM]].

%\cite{Kerr:2020qdo}
\bibitem{Kerr:2020qdo}
M.~Kerr \textit{et al.},
``The Parkes Pulsar Timing Array project: second data release,''
Publ. Astron. Soc. Austral. \textbf{37}, e020 (2020).

%\cite{Desvignes:2016yex}
\bibitem{Desvignes:2016yex}
G.~Desvignes \textit{et al.},
``High-precision timing of 42 millisecond pulsars with the European Pulsar Timing Array,''
Mon. Not. Roy. Astron. Soc. \textbf{458}, no.3, 3341-3380 (2016).

%\cite{Perera:2019sca}
\bibitem{Perera:2019sca}
B.~B.~P.~Perera, \textit{et al.},
``The International Pulsar Timing Array: Second data release,''
Mon. Not. Roy. Astron. Soc. \textbf{490}, no.4, 4666-4687 (2019).

%\cite{Kosowsky:1992rz}
\bibitem{Kosowsky:1992rz}
A.~Kosowsky, M.~S.~Turner and R.~Watkins,
``Gravitational waves from first order cosmological phase transitions,''
Phys. Rev. Lett. \textbf{69}, 2026-2029 (1992).

%\cite{Caprini:2010xv}
\bibitem{Caprini:2010xv}
C.~Caprini, R.~Durrer and X.~Siemens,
``Detection of gravitational waves from the QCD phase transition with pulsar timing arrays,''
Phys. Rev. D \textbf{82}, 063511 (2010).


%\cite{Nakai:2020oit}
\bibitem{Nakai:2020oit}
Y.~Nakai, M.~Suzuki, F.~Takahashi and M.~Yamada,
``Gravitational Waves and Dark Radiation from Dark Phase Transition: Connecting NANOGrav Pulsar Timing Data and Hubble Tension,''
Phys. Lett. B \textbf{816}, 136238 (2021).

%\cite{Addazi:2020zcj}
\bibitem{Addazi:2020zcj}
A.~Addazi, Y.~F.~Cai, Q.~Gan, A.~Marciano and K.~Zeng,
``NANOGrav results and dark first order phase transitions,''
Sci. China Phys. Mech. Astron. \textbf{64}, no.9, 290411 (2021).

%\cite{Ratzinger:2020koh}
\bibitem{Ratzinger:2020koh}
W.~Ratzinger and P.~Schwaller,
``Whispers from the dark side: Confronting light new physics with NANOGrav data,''
SciPost Phys. \textbf{10}, no.2, 047 (2021).

%\cite{Li:2021qer}
\bibitem{Li:2021qer}
S.~L.~Li, L.~Shao, P.~Wu and H.~Yu,
``NANOGrav signal from first-order confinement-deconfinement phase transition in different QCD-matter scenarios,''
Phys. Rev. D \textbf{104}, no.4, 043510 (2021).

%\cite{Machado:2018nqk}
\bibitem{Machado:2018nqk}
C.~S.~Machado, W.~Ratzinger, P.~Schwaller and B.~A.~Stefanek,
``Audible Axions,''
JHEP \textbf{01}, 053 (2019).

%\cite{Machado:2019xuc}
\bibitem{Machado:2019xuc}
C.~S.~Machado, W.~Ratzinger, P.~Schwaller and B.~A.~Stefanek,
``Gravitational wave probes of axionlike particles,''
Phys. Rev. D \textbf{102}, no.7, 075033 (2020).

%\cite{Salehian:2020dsf}
\bibitem{Salehian:2020dsf}
B.~Salehian, M.~A.~Gorji, S.~Mukohyama and H.~Firouzjahi,
``Analytic study of dark photon and gravitational wave production from axion,''
JHEP \textbf{05}, 043 (2021).

%\cite{Hiramatsu:2013qaa}
\bibitem{Hiramatsu:2013qaa}
T.~Hiramatsu, M.~Kawasaki and K.~Saikawa,
``On the estimation of gravitational wave spectrum from cosmic domain walls,''
JCAP \textbf{02}, 031 (2014).

%\cite{Kadota:2015dza}
\bibitem{Kadota:2015dza}
K.~Kadota, M.~Kawasaki and K.~Saikawa,
``Gravitational waves from domain walls in the next-to-minimal supersymmetric standard model,''
JCAP \textbf{10}, 041 (2015).


%\cite{Siemens:2006yp}
\bibitem{Siemens:2006yp}
X.~Siemens, V.~Mandic and J.~Creighton,
``Gravitational wave stochastic background from cosmic (super)strings,''
Phys. Rev. Lett. \textbf{98}, 111101 (2007).

%\cite{Blanco-Pillado:2017rnf}
\bibitem{Blanco-Pillado:2017rnf}
J.~J.~Blanco-Pillado, K.~D.~Olum and X.~Siemens,
``New limits on cosmic strings from gravitational wave observation,''
Phys. Lett. B \textbf{778}, 392-396 (2018).

%\cite{Ellis:2020ena}
\bibitem{Ellis:2020ena}
J.~Ellis and M.~Lewicki,
``Cosmic String Interpretation of NANOGrav Pulsar Timing Data,''
Phys. Rev. Lett. \textbf{126}, no.4, 041304 (2021).

%\cite{Blasi:2020mfx}
\bibitem{Blasi:2020mfx}
S.~Blasi, V.~Brdar and K.~Schmitz,
``Has NANOGrav found first evidence for cosmic strings?,''
Phys. Rev. Lett. \textbf{126}, no.4, 041305 (2021).

%\cite{Vaskonen:2020lbd}
\bibitem{Vaskonen:2020lbd}
V.~Vaskonen and H.~Veerm\"ae,
``Did NANOGrav see a signal from primordial black hole formation?,''
Phys. Rev. Lett. \textbf{126}, no.5, 051303 (2021).

%\cite{DeLuca:2020agl}
\bibitem{DeLuca:2020agl}
V.~De Luca, G.~Franciolini and A.~Riotto,
``NANOGrav Data Hints at Primordial Black Holes as Dark Matter,''
Phys. Rev. Lett. \textbf{126}, no.4, 041303 (2021).

%\cite{Kohri:2020qqd}
\bibitem{Kohri:2020qqd}
K.~Kohri and T.~Terada,
``Solar-Mass Primordial Black Holes Explain NANOGrav Hint of Gravitational Waves,''
Phys. Lett. B \textbf{813}, 136040 (2021).

%\cite{Press:1989yh}
\bibitem{Press:1989yh}
W.~H.~Press, B.~S.~Ryden and D.~N.~Spergel,
``Dynamical Evolution of Domain Walls in an Expanding Universe,''
Astrophys. J. \textbf{347}, 590-604 (1989).

%\cite{Garagounis:2002kt}
\bibitem{Garagounis:2002kt}
T.~Garagounis and M.~Hindmarsh,
``Scaling in numerical simulations of domain walls,''
Phys. Rev. D \textbf{68}, 103506 (2003).

%\cite{Leite:2011sc}
\bibitem{Leite:2011sc}
A.~M.~M.~Leite and C.~J.~A.~P.~Martins,
``Scaling Properties of Domain Wall Networks,''
Phys. Rev. D \textbf{84}, 103523 (2011).

%\cite{Leite:2012vn}
\bibitem{Leite:2012vn}
A.~M.~M.~Leite, C.~J.~A.~P.~Martins and E.~P.~S.~Shellard,
``Accurate Calibration of the Velocity-dependent One-scale Model for Domain Walls,''
Phys. Lett. B \textbf{718}, 740-744 (2013).

\bibitem{Mazumdar19}
A. Mazumdar and G. White, 
``Review of cosmic phase transitions: their significance and experimental signatures,''
Rep. Prog. Phys. \textbf{82}, 076901 (2019).

\bibitem{Kibble1976}
T. Kibble, ``Topology of cosmic domains and strings,''
J. Phys. A \textbf{9}, 1387 (1976).

%\cite{Jeannerot:2003qv}
\bibitem{Jeannerot:2003qv}
R.~Jeannerot, J.~Rocher and M.~Sakellariadou,
``How generic is cosmic string formation in SUSY GUTs,''
Phys. Rev. D \textbf{68}, 103514 (2003).



%\cite{Blanco-Pillado:2013qja}
\bibitem{Blanco-Pillado:2013qja}
J.~J.~Blanco-Pillado, K.~D.~Olum and B.~Shlaer,
``The number of cosmic string loops,''
Phys. Rev. D \textbf{89}, no.2, 023512 (2014).


\bibitem{Vilenkin1981}
A. Vilenkin, ``Gravitational radiation from cosmic strings,''
Phys. Lett. B \textbf{107}, 47 (1981).

\bibitem{Turok1984}
N. Turok, ``Grand unified strings and galaxy formation,''
Nucl. Phys. B \textbf{242}, 520 (1984).

\bibitem{Quashnock1990}
J. M. Quashnock and D. N. Spergel,
``Gravitational self-interactions of cosmic strings,''
Phys. Rev. D \textbf{42}, 2505 (1990).


%\cite{Cui:2017ufi}
\bibitem{Cui:2017ufi}
Y.~Cui, M.~Lewicki, D.~E.~Morrissey and J.~D.~Wells,
``Cosmic Archaeology with Gravitational Waves from Cosmic Strings,''
Phys. Rev. D \textbf{97}, no.12, 123505 (2018).

%\cite{Cui:2018rwi}
\bibitem{Cui:2018rwi}
Y.~Cui, M.~Lewicki, D.~E.~Morrissey and J.~D.~Wells,
``Probing the pre-BBN universe with gravitational waves from cosmic strings,''
JHEP \textbf{01}, 081 (2019).

%\cite{Blanco-Pillado:2019vcs}
\bibitem{Blanco-Pillado:2019vcs}
J.~J.~Blanco-Pillado, K.~D.~Olum and J.~M.~Wachter,
``Energy-conservation constraints on cosmic string loop production and distribution functions,''
Phys. Rev. D \textbf{100}, no.12, 123526 (2019).

%\cite{Blanco-Pillado:2019tbi}
\bibitem{Blanco-Pillado:2019tbi}
J.~J.~Blanco-Pillado and K.~D.~Olum,
``Direct determination of cosmic string loop density from simulations,''
Phys. Rev. D \textbf{101}, no.10, 103018 (2020).

\bibitem{IPTADR2}
https://zenodo.org/record/5787557


\bibitem{Hind1995}
M. Hindmarsh and T. Kibble, 
``Cosmic strings,''
Rep. Prog. Phys. \textbf{58}, 477 (1995).


%\cite{King:2020hyd}
\bibitem{King:2020hyd}
S.~F.~King, S.~Pascoli, J.~Turner and Y.~L.~Zhou,
``Gravitational Waves and Proton Decay: Complementary Windows into Grand Unified Theories,''
Phys. Rev. Lett. \textbf{126}, no.2, 021802 (2021).

%\cite{Trotta:2005ar}
\bibitem{Trotta:2005ar}
R.~Trotta,
``Applications of Bayesian model selection to cosmological parameters,''
Mon. Not. Roy. Astron. Soc. \textbf{378}, 72-82 (2007).

%\cite{Wang:2022wwj}
\bibitem{Wang:2022wwj}
D.~Wang,
``Squeezing Cosmological Phase Transitions with International Pulsar Timing Array,''
[arXiv:2201.09295 [astro-ph.CO]].

%\cite{JacksonKimball:2017elr}
\bibitem{JacksonKimball:2017elr}
D.~F.~J. Kimball \textit{et al.},
``Overview of the Cosmic Axion Spin Precession Experiment (CASPEr),''
Springer Proc. Phys. \textbf{245}, 105-121 (2020).
























































\end{thebibliography}
\end{document}